\documentclass[12pt,letterpaper]{article}
\usepackage{amsmath}
\usepackage{amssymb}
\usepackage{latexsym}
\usepackage{hyperref}
\usepackage{enumitem}
\newcommand{\be}{\begin{equation}}
\newcommand{\ee}{\end{equation}}
\newcommand{\bea}{\begin{eqnarray}}
\newcommand{\eea}{\end{eqnarray}}

\begin{document} 

\title{Beyond Falsifiability: \\Normal Science in a Multiverse}

 \author{Sean M. Carroll\\
 Walter Burke Institute for Theoretical Physics\\
 California Institute of Technology, Pasadena, CA 91125, U.S.A.\\
 seancarroll@gmail.com}
\maketitle

\begin{quotation} \noindent
Cosmological models that invoke a multiverse -- a collection of unobservable regions of space where conditions are very different from the region around us -- are controversial, on the grounds that unobservable phenomena shouldn't play a crucial role in legitimate scientific theories.
I argue that the way we evaluate multiverse models is precisely the same as the way we evaluate any other models, on the basis of abduction, Bayesian inference, and empirical success.
There is no scientifically respectable way to do cosmology without taking into account different possibilities for what the universe might be like outside our horizon.
Multiverse theories are utterly conventionally scientific, even if evaluating them can be difficult in practice.
\end{quotation}


\vfill
\noindent
Invited contribution to \textit{Epistemology of Fundamental Physics: Why Trust a Theory?}, eds. R. Dawid, R. Dardashti, and K. Th\'ebault (Cambridge University Press). CALT 2018-003.

\newpage

\section{Introduction}

The universe seems to be larger than the part we can see.
The Big Bang happened a finite time ago, and light moves at a finite speed, so there is only a finite amount of space accessible to our observations.
In principle space could be a compact manifold with a size comparable to that of our observable region, but there is no evidence for such a possibility, and it seems likely that space extends much further than what we see \cite{Vaudrevange:2012da}.
It is natural, then, for scientists to wonder what the universe is like beyond our observable part of it.

Or is it? 

There are two basic options for the ultra-large-scale universe that are worth distinguishing.
The first is uniformity.
The universe we see is approximately homogeneous and isotropic on large scales.
A simple possibility is that this uniformity extends throughout all of space, which might itself be finite or infinite. 
But another possibility is that conditions beyond our horizon eventually start looking very different.
At a prosaic level, the average density of matter, or the relative abundances of ordinary matter and dark matter, could vary from place to place.
More dramatically, but consistent with many ideas about fundamental physics, even the local laws of physics could appear different in different regions -- the number and masses of particles, forms and strengths of interactions between them, or even the number of macroscopic dimensions of space.
The case where there exist many such regions of space, with noticeably different local conditions, has been dubbed the ``cosmological multiverse,'' even if space itself is connected. (It would be equally valid but less poetic to simply say ``a universe where things are very different from place to place on ultra-large scales.")

For a number of reasons, both cosmologists and non-cosmologists have paid a lot of attention to the prospect of a cosmological multiverse in recent years \cite{Weinberg:2005fh,kragh}.
At the same time, however, there has been a backlash; a number of highly respected scientists have objected strongly to the idea, in large part due to a conviction that what happens outside the universe we can possibly observe simply shouldn't \textit{matter} \cite{es,smolin,isl,ellis}.
The job of science, in this view, is to account for what we observe, not to speculate about what we don't.
There is a real worry that the multiverse represents imagination allowed to roam unfettered from empirical observation, unable to be tested by conventional means.

In its strongest from, the objection argues that the very idea of an unobservable multiverse shouldn't count as science at all, often appealing to Karl Popper's dictum that a theory should be falsifiable to be considered scientific.
At the same time, proponents of the multiverse (and its partner in crime, the anthropic principle) will sometimes argue that while multiverse cosmologies are definitely part of science, they represent a new \textit{kind} of science, ``a deep change of paradigm that revolutionizes our understanding of nature and opens new fields of possible scientific thought" \cite{barrau}.
(Similar conceptual issues are relevant for other kinds of theories, including string theory and the many-worlds of Everettian quantum mechanics, but to keep things focused here I will only discuss the cosmological multiverse.)

In this essay I will stake out a judicious middle position.
Multiverse models are scientific in an utterly conventional sense; they describe definite physical situations, and are ultimately judged on their ability to provide an explanation for data collected in observations and experiments.
But the kind of science they are is perfectly ordinary science.
The ways in which we evaluate the multiverse as a scientific hypothesis are precisely the ways in which hypotheses have always been judged (cf. \cite{harlow}).

The point is not that we are changing the nature of science by allowing unfalsifiable hypotheses into our purview.
The point is that ``falsifiability'' was never the way that scientific theories were judged (although scientists have often talked as if it were).
While the multiverse is a standard scientific hypothesis, it does highlight interesting and nontrivial issues in the methodology and epistemology of science.
The best outcome of current controversies over the multiverse and related ideas (other than the hopeful prospect of finding the correct description of nature) is if working scientists are nudged toward accepting a somewhat more nuanced and accurate picture of scientific practice.

\section{Falsifiability and Its Discontents}

Falsifiability arises as an answer to the ``demarcation problem'' -- what is science and what is not?
The criterion, introduced by Karl Popper \cite{popper1}, states that truly scientific theories are ones that stick their necks out, offering specific predictions we might imagine testing by ``possible or conceivable observations" \cite{popper2}.
If a theory could in principle be proven false by the appropriate experimental results, it qualifies as scientific; if not, it doesn't.
Indeed, the hope was that such a principle could fully capture the logic of scientific discovery: we imagine all possible falsifiable theories, set about falsifying all the ones that turn out to be false, and what remains standing will be the truth.

Popper was offering an alternative to the intuitive idea that we garner support for ideas by verifying or confirming them.
In particular, he was concerned that theories such as the psychoanalysis of Freud and Adler, or Marxist historical analysis, made no definite predictions; no matter what evidence was obtained from patients or from history, one could come up with a story within the appropriate theory that seemed to fit all of the evidence.
Falsifiability was meant as a corrective to the claims of such theories to scientific status.

On the face of it, the case of the multiverse seems quite different than the theories Popper was directly concerned with.
There is no doubt that any particular multiverse scenario makes very definite claims about what is true.
Such claims could conceivably be falsified, if we allow ourselves to count as ``conceivable'' observations made outside our light cone. (We can't actually make such observations in practice, but we can conceive of them.)
So whatever one's stance toward the multiverse, its potential problems are of a different sort than those raised (in Popper's view) by psychoanalysis or Marxist history.

More broadly, falsifiability doesn't actually work as a solution to the demarcation problem, for reasons that have been discussed at great length by philosophers of science.
In contrast with the romantic ideal of a killer experiment that shuts the door on a theory's prospects, most scientific discoveries have a more ambiguous impact, especially at first. 
When the motions of the outer planets were found to be anomalous, nobody declared that Newtonian gravity had been falsified; they posited the existence of a new planet, and eventually discovered Neptune.
When a similar thing happened with Mercury, they tried the same trick, but this time the correct conclusion was entirely different; Newtonian gravity actually had been falsified, as we understood once general relativity came on the scene. 
When Einstein realized that general relativity predicted an expanding or contracting universe, he felt free to alter his theory by adding a cosmological constant, to bring it in line with the understanding at the time that the universe was static.
When the OPERA experiment discovered neutrinos that apparently traveled faster than light, very few physicists thought that special relativity had been falsified.
Individual experiments are rarely probative.

In any of these cases, it's simple to explain why an overly naive version of falsifiability failed to capture the reality of the scientific process: experimental results can be wrong or misinterpreted, theories can be tweaked to fit the data.
This is the point.
Science proceeds via an ongoing dialogue between theory and experiment, searching for the best possible understanding, rather than cleanly lopping off falsified theories one by one.
(Popper himself thought Marxism had started out scientific, but had become unfalsifiable over time as its predictions failed to come true.)

While philosophers of science have long since moved past falsifiability as a simple solution to the demarcation problem, many scientists have seized on it with gusto, going so far as to argue that falsifiability is manifestly a central part of the definition of science.
In the words of philosopher Alex Broadbent,
\begin{quote}
It is remarkable and interesting that Popper remains extremely popular among natural scientists, despite almost universal agreement among philosophers that -- notwithstanding his ingenuity and philosophical prowess -- his central claims are false.  \cite{broadbent}
\end{quote}
It is precisely challenging cases such as the multiverse that require us to be a bit more careful and accurate in our understanding of the scientific process.

Our goal here, however, is not to look into what precisely Karl Popper might have said or thought.
Even if the most naive reading of the falsifiability criterion doesn't successfully separate science from non-science, it does seem to get at something important about the nature of a scientific theory.
If we want to have a defensible position on how to view the multiverse hypothesis, it would be helpful to figure out what is useful about falsifiability, and how the essence of the idea can be adapted to realistic situations.

Falsifiability gets at two of the (potentially many) important aspects of science:
\begin{itemize}
\item \textbf{Definiteness.} A good scientific theory says something specific and inflexible about how nature works. It shouldn't be possible to take any conceivable set of facts and claim that they are compatible with the theory.
\item \textbf{Empiricism.} The ultimate purpose of a theory is to account for what we observe. No theory should be judged to be correct solely on the basis of its beauty or reasonableness or other qualities that can be established without getting out of our armchairs and looking at the world.
\end{itemize}
The falsifiability criterion attempts to operationalize these characteristics by insisting that a theory say something definite that can then be experimentally tested.
The multiverse is an example of a theory that says something definite (there are other regions of space far away where conditions are very different), but in such a way that this prediction cannot be directly tested, now or at any time in the future.

But there is a crucial difference between ``makes a specific prediction, but one we can't observe'' and ``has no explanatory impact on what we do observe."
Even if the multiverse itself is unobservable, its existence may well change how we account for features of the universe we \emph{can} observe.
Allowing for such a situation is an important way in which the most naive imposition of a falsifiability criterion would fail to do justice to the reality of science as it is practiced.

To try to characterize what falsifiability gets at, we can distinguish between different ways of construing what it might mean for a theory to be falsifiable.
Levels of falsifiability, from least to most, might include:
\begin{enumerate}
\item There is no conceivable empirical test, in principle or in practice or in our imaginations, that could return a result that is  incompatible with the theory. This would be Popper's view of the theories of Freud, Adler, and Marx.
\item There exist tests that are imaginable, but it is impossible for us to ever do them. This is the situation we could be in with the multiverse, where we can conceive of looking beyond our cosmological horizon, but can't actually do so.
\item There exist tests that are possible to do within the laws of physics, but are hopelessly impractical. We can contemplate building a particle accelerator the size of our galaxy, but it's not something that will happen no matter how far technology progresses; likewise, we can imagine decoherent branches of an Everettian wave function recohering, but the timescales over which that becomes likely are much longer than the age of the universe.
\item There exist tests that can be performed, but which will only ever cover a certain subset of parameter space for the theory. Models of dynamical dark energy (as opposed to a non-dynamical cosmological constant) can be parameterized by the rate at which the energy density evolves; this rate can be constrained closer and closer to zero, but it's always conceivable that the actual variation is below any particular observational limit yet doesn't strictly vanish.
\item There exist doable, definitive tests that could falsify a theory. General relativity would be falsified if large masses were shown not to deflect passing light.
\end{enumerate}

Most scientists would probably agree that theories in category 1.\ aren't really scientific in any practical sense, while those in category 5.\ certainly are (at least as far as testability is concerned). 
But the purpose of the list isn't to figure out where to definitively draw the line between ``scientific'' and ``non-scientific" theories; for the purposes of scientific practice, in any event, it is not clear what use such a distinction actually has.
Nevertheless, this list helps us to appreciate the sharp divide between category 1.\ and categories 2.-5. 
That difference can be summarized by pointing out that any theory in categories 2.-5.\ might reasonably be \emph{true} -- it could provide a correct description of nature, to some degree of accuracy and within some domain of applicability.
Theories in category 1., by contrast, don't really even have a chance of being true; they don't successfully distinguish true things from not-true things, regardless of their testability by empirical means.
Such theories are unambiguously not helpful to the scientific enterprise.

It would be wrong, therefore, to lump together categories 1.-4.\  in opposition to 5., while glossing over the distinction between 1.\ and 2.-5.
The former distinction is epistemological: it is a statement about how hard it is to figure out whether what the theories are saying is actually true.
The latter distinction is metaphysical: there is a real sense in which theories in category 1.\ aren't actually saying anything about the world.

The multiverse as a general paradigm -- regions far outside our observable part of the universe, where local conditions or even low-energy laws of physics could be very different -- is in category 2.
The universe could be like that, and there is nothing unscientific about admitting so.
The fact that it might be difficult to gather empirical evidence for or against such a possibility doesn't change the fact that it might be true.

Another way to emphasize that the important dividing line is between categories 1.\ and 2., not between 4.\ and 5., is to note that some specific multiverse models are actually in category 4., where a certain region of parameter space allows the theory to be explicitly falsified.
For the moment let's put aside the general idea of ``the universe looking very different outside our observable region," and focus on the most popular realization of this idea in contemporary cosmology, known as ``false-vacuum eternal inflation" \cite{Kleban:2011pg}.
In this kind of scenario, bubbles of lower-energy configurations appear via quantum tunneling within a region of space undergoing inflation; these bubbles grow at nearly the speed of light, and can contain within them distinct ``universes'' with potentially different local laws of physics.
The distribution of bubbles depends on details of physics at high energies -- details about which we currently have next to no firm ideas.
But we can parameterize our ignorance in terms the nucleation rate of such bubbles and the energy density within them.
If the rate is sufficiently high, this model makes a falsifiable prediction: the existence of circular features in the anisotropy of the cosmic microwave background, remnants of when other bubbles literally bumped into our own.
Cosmologists have searched for such a signature, allowing us to put quantitative limits on the parameters of false-vacuum eternal inflation models \cite{Kleban:2011yc,Braden:2016tjn}.

This particular version of the multiverse, in other words, is indubitably falsifiable for certain parameter values.
It would be strange indeed if the status of an idea as scientific vs. unscientific were parameter-dependent.
One might suggest that models with such tunable parameters should be labeled unscientific for \emph{any} parameter values -- even the testable ones -- on the grounds that such theories can wriggle out of being falsified by clever choice of parameters.
But in actual scientific practice that's not how we behave.
In the dark-energy example given above, the density of dark matter is imagined to vary at some slow rate; if the rate is slow enough, we will never be able to distinguish it from a truly constant vacuum energy.
Nobody claims that therefore the idea of dynamical dark energy is unscientific, nor should they.
Instead, we test the theory where we can, and our interest in the idea gradually declines as it is squeezed into ever-small regions of parameter space.

In the end, ``what counts as science" is less important than how science gets done.
Science is about what is true.
Any proposal about how the world truly is should be able to be judged by the methods of science, even if the answer is ``we don't know and never will."
Excluding a possible way the world could be on the basis of a philosophical predisposition is contrary to the spirit of science.
The real question is, how should scientific practice accommodate this possibility?

\section{Abduction and Inference}

While we don't currently have a final and complete understanding of how scientific theories are evaluated, we can do a little bit better than the falsifiability criterion.
The multiverse provides an especially interesting test case for attempts to explain how science progresses in the real world.

One way of thinking about the scientific method is as an example of ``abduction,'' or inference to the best explanation, which is to be contrasted with the logical techniques of deduction and induction \cite{sep-abduction}. 
Deduction reasons perfectly from a given set of axioms; induction attempts to discern patterns in repeated observations.
Abduction, by contrast, takes a given set of data and attempts to find the most likely explanation for them.
If you eat spoiled food, and you know that eating spoiled food usually makes you sick, by deduction you can conclude that you are likely to get sick.
Abduction lets you go the other way around: if you get sick, and there are no other obvious reasons why, you might conclude that you probably ate some spoiled food.
This isn't a logically necessary conclusion, but it's the best explanation given the context.
Science works along analogous lines.

This raises the knotty question of what makes an explanation ``the best."
Thomas Kuhn, after his influential book \textit{The Structure of Scientific Revolutions} led many people to think of him as a relativist when it came to scientific claims, attempted to correct this misimpression by offering a list of criteria that scientists use in practice to judge one theory better than another one: accuracy, consistency, broad scope, simplicity, and fruitfulness \cite{kuhn}.
``Accuracy'' (fitting the data) is one of these criteria, but by no means the sole one.
Any working scientist can think of cases where each of these concepts has been invoked in favor of one theory or another.
But there is no unambiguous algorithm according to which we can feed in these criteria, a list of theories, and a set of data, and expect the best theory to pop out.
The way in which we judge scientific theories is inescapably reflective, messy, and human.
That's the reality of how science is actually done; it's a matter of judgment, not of drawing bright lines between truth and falsity or science and non-science.
Fortunately, in typical cases the accumulation of evidence eventually leaves only one viable theory in the eyes of most reasonable observers.

This kind of abductive reasoning can be cast in quantitative terms by turning to Bayesian inference.
In this procedure, we imagine an exhaustive set of theories $T_i$ (one of which might be a catch-all ``some theory we haven't thought of yet"), and to each we assign a prior probability that the theory is true, $P(T_i)$. 
Then we collect some new data, $D$. 
Bayes's Theorem tells us $P(T_i|D)$, the posterior probability that we should assign to each theory in light of the new data, via the formula
\be
  P(T_i|D) = \frac{P(D|T_i) P(T_i)}{P(D)}.
\ee
Here, $P(D|T_i)$ is the likelihood of obtaining that data if the theory were true, and the normalization factor $P(D) = \sum_i P(D|T_i)P(T_i)$ is simply the total probability of obtaining the data under the weighted set of all theories.

Bayes's Theorem is a mathematical theorem; it is necessarily true under the appropriate assumptions.
The additional substantive claim is that this theorem provides the correct model for how we think about our credences is scientific theories, and how we update those credences in light of new information.

The role of priors is crucial.
Our priors depend on considerations such as Kuhn's criteria (aside from ``accuracy,'' which is a posterior notion).
Consider two theories of gravitation.
Theory One is precisely Einstein's general relativity.
Theory Two is also general relativity, but with the additional stipulation that Newton's constant will suddenly change sign in the year 2100, so that gravity becomes repulsive rather than attractive.\footnote{Any similarities to Nelson Goodman's invented property ``grue,'' meaning ``green'' before some specified future date and ``blue'' afterwards \cite{goodman}, are entirely intentional.}
Every sensible scientist will assign different credences to these two theories, even though the empirical support for them is exactly the same and they are equally falsifiable.
The second theory would be considered extremely unlikely, as it is unnecessarily complicated without gaining any increase in consistency, scope, or fruitfulness over conventional general relativity.

How does the shift from an overly simplistic ``falsification'' paradigm to one based on abduction and Bayesian inference affect our attitude toward the multiverse?
Rather than simply pointing out that the multiverse cannot be directly observed (at least for some parameter values) and therefore can't be falsified and therefore isn't science, we should ask whether a multiverse scenario might provide the best explanation for the data we do observe, and attempt to quantify this by assigning priors to different possibilities then updating them in the face of new data -- just as we do for any other scientific theory.

Put this way, it becomes clear that there isn't that much qualitatively special about the multiverse hypothesis.
It's an idea about nature that could be true or false, and to which we can assign some prior probability, which then gets updated as new information comes in.

Indeed, this is precisely what happened in 1998, when observations indicated that the universe is accelerating, something that is most easily explained by the existence of a nonzero cosmological constant (vacuum energy) \cite{Riess:1998cb,Perlmutter:1998np,Carroll:2000fy}.
Before this discovery was made, it was known that the cosmological constant had to be much smaller than its supposedly-natural value at the Planck scale, but it wasn't known whether it was exactly zero or just very close. (For convenience I am not distinguishing here between a truly constant vacuum energy and some kind of dynamical dark energy.)
At that point we could distinguish between four broad theories of the cosmological constant (each of which might include multiple more specific theories): 
\begin{enumerate} [label=\arabic*)]
\item There is a mechanism that sets the cosmological constant equal to zero once and for all.
\item There is a mechanism that sets the cosmological constant to some small nonzero value.
\item There is no mechanism that sets the cosmological constant to a small number, but it nevertheless is just by random chance. 
\item The cosmological constant takes on different values in different parts of a multiverse, and is small in our observable region for anthropic reasons. (If the vacuum energy is too large and positive, galaxies never form; if it is too large and negative, the universe rapidly recollapses. The vacuum energy must be small in magnitude relative to the Planck scale to allow for the existence of life.)
\end{enumerate}

Within the community of theoretical physicists, there was (and is) very little support for option 3), and pre-1998 a substantial prior was attached to option 1).
The reasoning why 1) was given a higher probability than 2) was essentially that it's hard to think of any mechanism that makes the vacuum energy very small, but in the set of all not-yet-proposed mechanisms that might someday be invented, it's easier to conceive of dynamics that set the cosmological constant all the way to zero (perhaps due to some unknown symmetry) than it is to imagine a mechanism that puts it just below the threshold of what had then been observed, but no lower.
In a 1987 article, Steven Weinberg \cite{Weinberg:1987dv} argued that under the multiverse scenario 4), we should actually expect the cosmological constant to be nonzero and near the observational bounds that existed at that time.
In other words, the likelihood of observing a nonzero cosmological constant was relatively large under 4) [and presumably under 3), though that was thought to be unlikely from the start], somewhat smaller under 2) [since a dynamical mechanism might prefer numbers very close to zero], and would be vanishing under 1).

Ten years later, when the evidence came in for an accelerating universe, scientists quite rightly updated the probabilities they attached to these theoretical alternatives.
By Bayes's Theorem, finding a nonzero cosmological constant decreased their credence in the popular theory 1), while increasing it in the other options.
In particular, this experimental result led directly to an increase in the credence assigned by working physicists to the multiverse hypothesis 4) \cite{Polchinski:2017vik}.
We didn't observe other universes directly, but we observed something about our universe that boosted the likelihood that a multiverse provides the best explanation for what we see.

The multiverse, therefore, is a case of science as usual: we evaluate it on the basis of how likely it is to be true, given what we know on the basis of what we actually have observed.
But it is not only examples of literal new data that can cause our credences in a theory to change.
The multiverse hypothesis reminds us of how better understanding, as well as actual experimental or observational input, can serve as ``data'' for the purposes of Bayesian inference.

Dawid, writing in the context of string theory, has highlighted the role of considerations other than simply fitting the data in evaluating theories \cite{dawid}.
He uses the term ``non-empirical confirmation,'' which has led to some marketing problems in the scientific community.
In epistemology, ``confirmation'' is used to refer to any evidence or considerations that increase our credence in a theory; it does not correspond exactly the the everyday-language notion of ``proving the truth of."
Even ``non-empirical'' is a bit misleading, as such considerations are only ever used in addition to empirical data, not in place of it.

But it is obviously true that our credences in scientific theories are affected by things other than collecting new data via observation and experiment.
Choosing priors is one obvious example, but an important role is also played by simply understanding a theory better (which is the job of theoretical physicists).
When Einstein calculated the precession of Mercury in general relativity and found that it correctly accounted for the known discrepancy with Newtonian gravity, there is no doubt that his credence in his theory increased substantially, as it should have.
No new \emph{data} were collected; the situtation concerning Mercury's orbit was already known.
But an improved theoretical understanding changed the degree of belief it made sense to have in that particular model.
When Gerard 't~Hooft showed that gauge theories with spontaneous symmetry breaking were renormalizable \cite{tHooft:1971qjg}, the credence in such models increased dramatically in the minds of high-energy physicists, and justifiably so.
Again, this change was not in response to new data, but to a better understanding of the theory.

These are sharp and relatively uncontroversial examples of credences changing as a result of improved theoretical understanding, but most changes are more gradual.
In the case of the multiverse, physicists have been greatly influenced by the realization that a multiverse need not be simply posited as a logical possibility, but is actually a \textit{prediction} of theories that are attractive for entirely different reasons: inflation and string theory.

It was realized early in the development of inflationary cosmology that inflation is often (depending on the model) eternal: once begun, inflation ends in some regions but continues in others, leading to the creation of different regions of space with potentially very different physical conditions \cite{steinhardt-nuffield,vilenkin-eternal,Guth:2007ng}.
String theory, meanwhile, was originally hoped to give a unique set of predictions for observable quantities, but more recently appears to be able to support an extremely large number of metastable vacuum states, each of which corresponds to different low-energy physics \cite{Bousso:2000xa,Kachru:2003aw,Susskind:2003kw}. (This realization was actually spurred in part by the discovery of a nonzero cosmological constant -- a rare instance where progress in string theory was catalyzed by experimental data.)
Taken together, these theories seem to predict that a multiverse is quite likely; correspondingly, anyone who assigns substantial credences to inflation and string theory would naturally assign substantial credence to the multiverse.

Once again, this is an example of science proceeding as usual.
Ideas are not evaluated in a vacuum; we ask how they fit in with other ideas we have reason to believe might play a role in the ultimate correct theory of the universe.
What theories predict, and how one theory fits together with another one, are perfectly respectable considerations to take into account when we assign credences to them.
And an improved understanding of these issues qualifies as ``data'' in the context of Bayes's Theorem.

None of which is to say that this kind of data could serve as a \textit{replacement} for the good old-fashioned evidence we collect in telescopes and microscopes.
Our priors for different ideas are naturally affected by non-empirical ideas such as elegance and beauty, and those may even affect our likelihoods, if for example we realize that there is an unanticipated hidden beauty underlying an existing theory; but an ability to account for the data is always the overwhelmingly important quality a scientific theory must have.
Our credence in general relativity increased due to a pencil-and-paper calculation by Einstein, but the impact of that calculation rested on the fact that it helped account for something we observed about the world, the precession of Mercury's perihelion.
Many physicists increased their credence in the multiverse when they realized that it was predicted by inflation and string theory, but inflation and string theory are ultimately interesting because they purport to explain features of observational reality -- the large-scale structure of the cosmos, and the existence of gravitation and gauge theories.
A correct accounting for the multitude of influences that shape our credences concerning scientific hypotheses is in no sense a repudiation of empiricism; it is simply an acknowledgment of the way it works in the real world.

\section{The Unavoidable Unobservable}

The best reason for classifying the multiverse as a straightforwardly scientific theory is that we don't have any choice.
This is the case for any hypothesis that satisfies two criteria:
\begin{itemize}
\item It might be true.
\item Whether or not it is true affects how we understand what we observe.
\end{itemize}

Consider a relevant example: attempting to solve the cosmological constant problem (understanding why the observed vacuum energy is so much smaller than the Planckian value we might naively expect).
Given that we don't currently know the answer, it might be the case that there exists some dynamical mechanism still waiting to be found, which will provide a simple formula that uniquely relates the observed cosmological constant to other measurable constants of nature.
Or it might be the case that no such mechanism exists.
As a working theoretical physicist interested in solving this problem, one cannot spend equal amounts of research effort contemplating every possible idea, as the number of ideas is extremely large. 
We use our taste, lessons from experience, and what we know about the rest of physics to help guide us in hopefully productive directions.

In this case, whether or not there is a multiverse in which the local cosmological constant takes on different values in different regions is an inescapable consideration.
If there is, then the need to find a unique dynamical mechanism is severely diminished, since no such mechanism can seemingly exist (given that the vacuum energy is different from place to place) and that there is a plausible alternative (anthropic selection).
If there is no multiverse, the interest in a unique dynamical mechanism is correspondingly enhanced.
Since our credence in the multiverse should never be exactly 0 or 1, individual scientists will naturally have different predilections toward pursuing different theoretical approaches, which is a sign of a healthy scientific community.
But it makes no sense to say that the possible existence of a multiverse is irrelevant to how we deal with the problem.

The universe beyond what we can see might continue uniformly forever, or it might feature a heterogeneous collection of regions with different local conditions.
Since direct observations cannot reveal which is the case, both alternatives are equally ``scientific."
To the extent that which alternative might be true affects our scientific practice, keeping both possibilities in mind is unavoidable.
The only unscientific move would be to reject one or the other hypotheses \textit{a priori} on the basis of an invented methodological principle.

None of which is to say that there aren't special challenges posed by the multiverse.
At a technical level, we have the measure problem: given an infinite multiverse, how do we calculate the relative probabilities of different local conditions?
Skeptics will sometimes say that since everything happens somewhere in the multiverse, it is impossible to make even probabilistic predictions.
Neither of these two clauses is necessarily correct; even if a multiverse is infinitely big, it does not follow that everything happens, and even if everything happens, it does not follow that there are no rules for the relative frequencies with which things happen.
After all, probabilistic theories such as statistical mechanics and quantum mechanics also require a measure, and in the case of cosmology several alternatives have been proposed  \cite{DeSimone:2008if,Freivogel:2011eg,Salem:2011qz}.
For inflation in particular, it seems sensible to imagine that the measure is strongly peaked on trajectories resembling traditional single-universe inflationary scenarios.
Given that eternal inflation is common in realistic inflationary potentials, some provisional assumption along these lines is necessary in order to license the interpretation of cosmological observables in terms of inflationary parameters (as most working cosmologists are more than willing to do).

There still remains the question, even if there is a correct measure on the multiverse, how will we ever \emph{know}? 
It seems hard to imagine doing experiments to provide an answer.
Rather, we are stuck with the more indirect kinds of reasoning mentioned above, especially how any particular proposal fits in with other physical theories.
(Presumably in this case that would involve some understanding of quantum gravity and the emergence of spacetime, which doesn't yet exist but might someday be forthcoming.)
And while such reasoning is necessary, it's hard for it to be definitive.
In Bayesian terms, we can nudge our credences upward or downward, but it is very hard on the basis of indirect evidence alone to send those credences so close to 0 or 1 that the question can be considered definitively answered.

That, in a nutshell, is the biggest challenge posed by the prospect of the multiverse.
It is not that the theory is unscientific, or that it is impossible to evaluate it.
It's that evaluating it is \emph{hard}.
It is entirely conceivable that we could make significant progress in understanding quantum gravity and inflationary cosmology, and yet we will be left with a situation where a century in the future our credence in the multiverse is still somewhere between 0.2 and 0.8. 

But the difficulty that we human beings might have in deciding whether a theory is correct should be of little relevance to the seriousness with which we contemplate it.
There really might be a multiverse out there, whether we like it or not.
To the extent that such a possibility affects how we attempt to explain features of the universe we do see, we need to treat this hypothesis just as we would any other scientific idea.
Nobody ever said science was going to be easy.

\section*{Acknowledgements}
 
This research is funded in part by the Walter Burke Institute for Theoretical Physics at Caltech and by DOE grant DE-SC0011632.

\end{document}